# Relation of the Weibull Shape Parameter with the Healthy Life Years Lost Estimates: Analytic Derivation and Estimation from an Extended Life Table

Christos H Skiadas[1] and Charilaos Skiadas[2]

[1] ManLab, Technical University of Crete, Chania, Crete, Greece
(E-mail: skiadas@cmsim.net )
[2] Department of Mathematics and Computer Science, Hanover College, Indiana, USA
(E-mail: skiadas@hanover.edu )

**Abstract**

Matsushita et al (1992) have done an interesting finding. They observed that the shape parameter of the Weibull model presented systematic changes over time and age when applied to mortality data for males and females in Japan. They have also estimated that this parameter was smaller in the 1891-1898 data in Japan compared to the 1980 mortality data and they presented an illustrative figure for females where the values of the shape parameter are illustrated on the diagram close to the corresponding survival curves. However, they have not provided an analytical explanation of this behavior of the shape parameter of the Weibull model. Of course the Weibull model is ideal to model the fatigue of materials. Especially the cumulative hazard of this model can express the additive process of applying a force for enough time before cracking. To pass to the human data, the Weibull model and the cumulative hazard can express the additive process which disabilities and diseases cause the human organism during the life span leading to healthy life years lost. In this paper we further analytically derive a more general model of survival-mortality in which we estimate a parameter related to the Healthy Life Years Lost (HLYL) and leading to the Weibull model and the corresponding shape parameter as a specific case. We have also demonstrated that the results found for the general HLYL parameter we have proposed provides results similar to those provided by the World Health Organization for the Healthy Life Expectancy (HALE) and the corresponding HLYL estimates. An analytic derivation of the mathematical formulas is presented along with an easy to apply Excel program. This program is an extension of the classical life table including four more columns to estimate the cumulative mortality, the average mortality, the person life years lost and finally the HLYL parameter $b_x$. The latest versions of this program appear in the Demographics2019 website at: http://www.asmda.es/demographics2019.html .

**The Weibull Model Revisited**

In this paper we explore a very interesting property of the Weibull model (Weibull, 1951) found last decades and published in few papers in books and journals (see Matsushita 1992, Skiadas and Skiadas (2014, 2015, 2018a,b,c, 2019), Skiadas and Arezzo, 2018, and Weong and Je 2011, 2012). Our findings had to do with the similarities of the shape parameter of this model with the estimated values of the Healthy

---

Life Years Lost (HLYL) provided by the World Health Organization (WHO). Both estimates from WHO and Weibull are close to each other for several countries. The next Table I summarizes several indicative cases for 2010 from European Countries with the higher Life Expectancy for males and females. The Table includes the Life Expectancy (LE), the Healthy Life Expectancy (HALE) as provided from the World Health Organization (WHO). Then we have calculated the Healthy Life Years Lost (HLYL) as the difference between the LE and the HALE to provide the WHO figures and our estimates for the Weibull shape parameter *b* and the Direct estimates (see Skiadas and Skiadas, 2018a,b,c, 2019) of the HLYL parameter $b_x$ from the Life Tables provided by the Human Mortality Database (HMD). At the last line of the Table the average values are provided. The Weibull and Direct estimates are similar for males and not very higher from the HLYL estimates from WHO. The standard error is 0.855 for the WHO vs Weibull comparison and 0.667 for the WHO vs Direct estimation. For females the Weibull and Direct estimates differ as the Weibull estimates with mean 11.28 are higher from the HLYL estimates from WHO (mean=10.32). The standard error is 1.107. The standard deviation is 0.437 for the WHO vs Direct estimate while the mean is 10.32 for WHO and 10.43 for the Direct estimation for females. In the following we introduce a quantitative methodology and mathematical analysis to explain this important finding and straighten further the applicability of this method in estimating the HLYL and the Healthy Life Expectancy.

TABLE I

Life Expectancy (LE), Healthy Life Expectancy (HALE), Healthy Life Years Lost (HLYL) from the World Health Organization (WHO) and our estimates for the Weibull shape parameter *b* and Direct estimates of the HLYL parameter $b_x$ from the Life Tables provided by the Human Mortality Database (HMD)

| | Males in 2010 | | | | | Females in 2010 | | | | |
|---|---|---|---|---|---|---|---|---|---|---|
| Country | LE | HALE | WHO | Weibull | Direct | LE | HALE | WHO | Weibull | Direct |
| Austria | 77.83 | 69.47 | 8.36 | 8.71 | 8.90 | 83.25 | 73.18 | 10.07 | 11.69 | 10.73 |
| Belgium | 77.53 | 69.15 | 8.38 | 8.75 | 8.78 | 82.70 | 72.34 | 10.36 | 11.09 | 10.31 |
| Czechia | 74.48 | 65.62 | 8.86 | 7.52 | 8.20 | 80.72 | 70.52 | 10.21 | 10.60 | 10.02 |
| Denmark | 77.35 | 69.17 | 8.18 | 8.55 | 8.48 | 81.39 | 71.64 | 9.75 | 9.60 | 9.15 |
| Finland | 76.76 | 68.05 | 8.71 | 8.34 | 8.73 | 83.24 | 72.64 | 10.60 | 11.64 | 10.69 |
| France | 78.20 | 70.19 | 8.02 | 8.56 | 8.84 | 84.52 | 74.06 | 10.46 | 11.74 | 10.68 |
| Germany | 77.57 | 69.30 | 8.27 | 8.67 | 8.77 | 82.60 | 72.48 | 10.13 | 11.34 | 10.52 |
| Greece | 77.98 | 69.84 | 8.14 | 8.58 | 8.52 | 83.12 | 73.05 | 10.08 | 11.75 | 10.79 |
| Ireland | 78.71 | 70.17 | 8.54 | 9.29 | 8.96 | 82.95 | 72.67 | 10.28 | 10.60 | 9.84 |
| Italy | 79.50 | 71.23 | 8.27 | 9.63 | 9.21 | 84.42 | 74.19 | 10.23 | 11.80 | 10.76 |
| Luxembourg | 78.80 | 70.01 | 8.80 | 9.21 | 9.02 | 83.70 | 73.14 | 10.56 | 11.43 | 10.61 |
| Netherlands | 78.81 | 70.36 | 8.45 | 9.40 | 9.08 | 82.73 | 72.42 | 10.31 | 10.98 | 10.31 |
| Norway | 78.88 | 70.30 | 8.58 | 9.54 | 9.51 | 83.15 | 73.23 | 9.92 | 11.09 | 10.42 |
| Portugal | 76.80 | 68.83 | 7.96 | 8.66 | 8.59 | 83.03 | 72.74 | 10.29 | 11.50 | 10.36 |
| Slovenia | 76.20 | 66.54 | 9.66 | 8.08 | 8.31 | 82.76 | 71.71 | 11.05 | 11.30 | 10.32 |
| Spain | 79.10 | 71.05 | 8.06 | 9.00 | 9.00 | 85.10 | 74.76 | 10.33 | 12.19 | 11.00 |
| Sweden | 79.55 | 70.83 | 8.72 | 9.86 | 9.50 | 83.48 | 72.97 | 10.51 | 11.42 | 10.71 |
| Switzerland | 80.08 | 71.22 | 8.86 | 9.96 | 9.69 | 84.51 | 73.65 | 10.87 | 12.30 | 11.25 |
| United Kingdom | 78.60 | 70.05 | 8.54 | 8.92 | 8.92 | 82.50 | 72.35 | 10.16 | 10.26 | 9.73 |
| **AVERAGE** | | | **8.49** | **8.91** | **8.90** | | | **10.32** | **11.28** | **10.43** |

**The HLYL Estimation Method**

Our methodology Skiadas and Skiadas (2014, 2015, 2018a,b,c, 2019) was based on a geometric approach based on the following graph of mortality spaces where both mortality and survival are presented as appropriate areas of this graph.

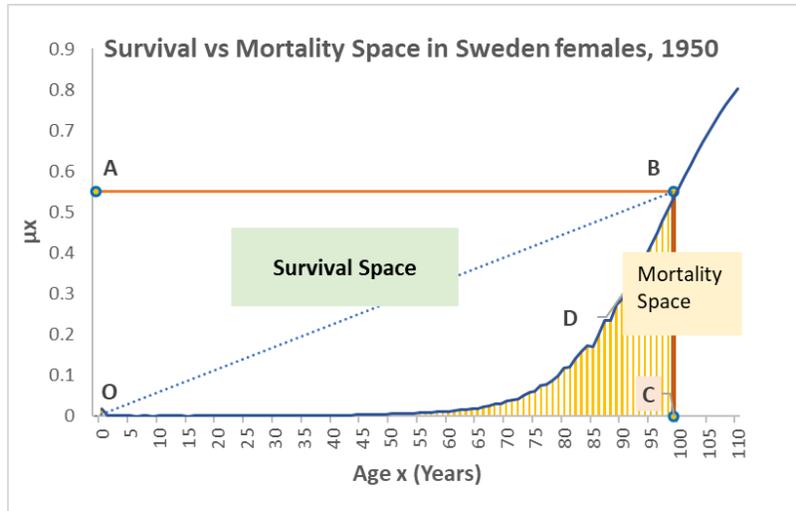

Fig. 1. Survival vs Mortality space graph

The usual form to express mortality $\mu_x$ in a population at age *x* is by estimating the fraction Death($D_x$)/Population($P_x$) that is $\mu_x=D_x/P_x$. As in the following we will use the Life Tables provided from the Human Mortality Database we will use the term *mx* of these tables instead of $\mu_x$. The above graph using data from Sweden 1950 females from the HMD is formulated with $\mu_x$ as the blue exponential curve. The main forms of Life Tables start with $\mu_x$ and in the following estimate the survival forms of the population. This methodology leads to the estimation of a probability measure termed as life expectancy at age *x* or life expectancy at birth when considering the total life time. There are several differences between the graph with the survival space above and the survival curves methodology. First of all, the vertical axis in the Survival-Mortality Space (SMS) diagram is the probability $\mu_x$. Instead in the survival diagram the vertical axis represent population (usually it starts from 100.000 in most life tables and gradually slow down until the end). By the SMS diagram we have probability spaces for both survival and mortality. For the age *x* the total space is (ABCOA) in the SMS diagrams that is (OA).(BC)=$x\,\mu_x$. The mortality space is the sum $S(\mu_x)$ and survival space is ($x\mu_x - S(\mu_x)$). Accordingly, the important measure of the Health State is simply the fraction (ABDOA)/(BCODB). Simpler is to prefer the fraction (ABCOA)/(BCODB)=$x\mu_x/S(\mu_x)$ that can be estimated from $\mu_x$ for every age *x* of the population.

Ruben Roman et al (2007) propose a similar methodology stating: "In the expression of the survival function; H(x) denotes the cumulative hazard function, which is equivalent to the area under the hazard function m(x). The area under the hazard function was defined by taking the corresponding integration limits ranging from x, current age of an individual, to x + $y_x$, age at death or quantity of time lived from birth to death, where X and $Y_x$ are non-negative continuous random variables. The calculated area will give the risk of dying at a given age x up to a particular future time $y_x$". The cumulative hazard they propose is $S\,\mu_x$ where $\mu_x$ is equivalent to the hazard function in our notation.

In modeling the healthy life years lost to disability some important issues should be realized. Mortality expressed by $\mu_x$ is important for modeling disability but more important is the cumulative mortality $S\,\mu_x$ which, as an additive process, is more convenient for the estimation of the healthy life years lived with disability and the deterioration process causing deaths. The estimates for this type of mortality are included in the term $b_x S\,\mu_x$.

Our approach in previous publications (Skiadas and Skiadas (2018a,b,c, 2019)) was to set a time varying fraction $b_x$ for Health/Mortality of the form:

$$b_x = \frac{x\mu_x}{\int_0^x \mu_s ds} \tag{1}$$

This formula is immediately provided from the last figure by considering the fraction:

$$b_x = \frac{Total\ Space}{Mortality\ Space} = \frac{OABCO}{ODBCO} = \frac{x\mu_x}{\int_0^x \mu_s ds}$$

It should be noted that an alternative approach is given by:

$$b_x = \frac{Survival\ Space}{Mortality\ Space} = \frac{OABCO - ODBCO}{ODBCO} = \frac{x\mu_x}{\int_0^x \mu_s ds} - 1$$

In the latter case the estimated fraction $b_x$ is smaller by one from the previous case. It remains to the applications stage to decide for the most appropriate. So far the Total Space approach is simpler and gave good results.

The main hypothesis is that the population involved in the deterioration process is a fraction of the total population determined by the level of mortality $\mu_x$ at age *x*. Accordingly the mortality process will have two alternatives expressed by the simple equation:

$$x\mu_x = b_x \int_0^x \mu_s ds \approx b_x \sum_0^x \mu_x \tag{2}$$

Where $x\mu_x$ is the incoming part related to the disability of the living population and the second part is the outgoing part that is summed to the mortality for the period from 0 to age *x*. The parameter $b_x$ is a corresponding adding to express the rate of healthy life lost to disability. The applications verify that the maximum values for $b=b_{max}$ are compatible to the estimates of the WHO for several countries. Evenmore, our estimates expressing the values for $b_x$ in all the life time are of particularly importance in the studies related to the Health Expenditure estimation.

Some important properties of the last formula are given below:

First we can formulate the Survival Probability *S(t)*

$$S(t) = \exp\left(-\int_0^x \mu_s ds\right) = \exp\left(-\frac{x\mu_x}{b_x}\right) \approx \exp\left(-\sum_0^x \mu_x\right) \tag{3}$$

$$S(t) \approx \exp\left(-\sum_0^x \mu_x\right) = \exp(-\mu_0)\exp(-\mu_1)\exp(-\mu_2)\ldots\exp(-\mu_x)$$

Next we can differentiate (2) to obtain

$$(x\mu_x)' = b'_x \int_0^x \mu_s ds + b_x \mu_x \tag{4}$$

For a constant $b$ we have $b'_x = 0$ and

$$x\mu'_x + \mu_x = b\mu_x$$

It follows

$$x\mu'_x = (b-1)\mu_x$$

And rearranging

$$\frac{\mu'_x}{\mu_x} = \frac{b-1}{x}$$

Solving the differential equation

$$\ln(\mu_x) = \ln(c) + (b-1)\ln x$$

Where c is a constant of integration. Finally

$$\mu_x = cx^{b-1}$$

By setting $c=\lambda b$ the hazard function or the generating function of the Weibull appear

$$\mu_x = \lambda b x^{b-1} \tag{5}$$

And the cumulative hazard of the Weibull is

$$\Lambda(x) = \lambda x^b = \frac{x\mu_x}{b} = \int_0^x \mu_s ds \tag{6}$$

This is to verify the formula for the survival probability (3) presented earlier.

Matsushita et al (1992) had suggested the Weibull model for a Lifetime Data Analysis of Disease and Aging. From the big variety of studies modeling and applying the Weibull model for more than 70 years this work is of particular importance as they emphasize to the introduction of the Weibull shape parameter in connection to survival rates. They have studied specific cases of Japanese females for 1891-1898 and for 1980. They have diagnosed the growing process of the shape parameter, *m* in their notation and *b* in our notation, during age and time as the maximum *m*=7.40 for the period 1891-1898 and m=9.19 for 1980. For the latter case our direct estimates give *b*=9.43. They have also presented similar changes of *m* for males and females for several age and time periods thus establishing a systematic variation of *m* over time and age. Evenmore, they have done calculations for *m* from data for several diseases. In the next Table II we have done estimates for the (Life Expectancy-Healthy Life Expectancy) = (LE-HALE) = HLYL for males and females in Japan. LE, HALE and LE-HALE are provided from the paper from Tokudome et al (2016) whereas the HLYL are estimated with our direct method. The estimates of the HLYL approach well between the two methodologies. The advantage with our methodology is that we can estimate the HLYL in all time periods as far as life table data exist. We have done projections for the Tokudome et al. estimates in figure 2 by fitting a line to the female data in Japan from 1990 to 2013. The estimates provided (LE-HALE)=9.19 precisely the same with the Matsushita et all. estimates.

| TABLE II | | | | | | | | |
|---|---|---|---|---|---|---|---|---|
| | Japan Males | | | | Japan Females | | | |
| | LE | HALE | LE-HALE | HLYL | LE | HALE | LE-HALE | HLYL |
| 1980 | | | | 7.9 | | | | 9.4 |
| 1985 | | | | 8.0 | | | | 9.8 |
| 1990 | 76.0 | 68.1 | 7.9 | 8.3 | 82.0 | 72.2 | 9.7 | 10.0 |
| 1995 | 76.5 | 68.4 | 8.1 | 8.4 | 82.2 | 72.9 | 9.9 | 10.1 |
| 2000 | 77.6 | 69.1 | 8.5 | 8.4 | 84.3 | 74.0 | 10.4 | 10.2 |
| 2005 | 78.7 | 69.9 | 8.8 | 8.6 | 85.5 | 74.8 | 10.7 | 10.5 |
| 2010 | 79.3 | 70.8 | 8.5 | 9.0 | 86.1 | 75.4 | 10.7 | 11.0 |
| 2013 | 80.1 | 71.1 | 8.9 | 9.1 | 86.4 | 75.6 | 10.8 | 11.1 |

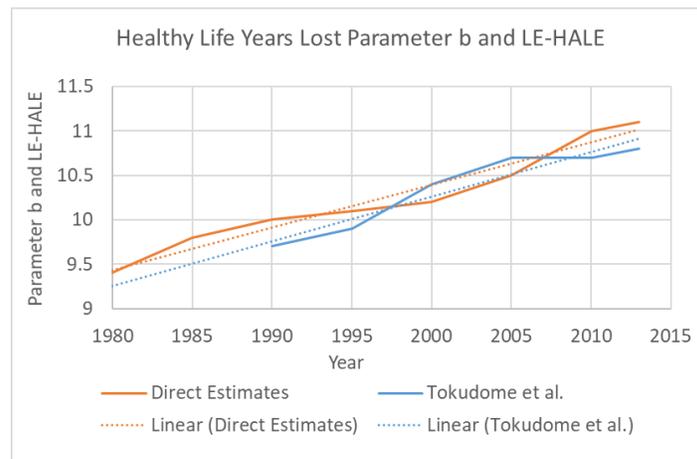

Fig. 2. Healthy Life Years Lost and extensions in Japan

As we already have presented in previous studies (Skiadas and Skiadas (2018a,b,c, 2019)), $b_x$ can be estimated directly from the life table data. By using this method, the resulting form is illustrated in the next figures. An almost study growth until a high level and then a decline at very high ages. Clearly this is not a simple equation form. However, there is a simple case to estimate $b_x$ at the top level by taking a small linear part of the $b_x$ curve at the top level parallel to the horizontal axis. This implies that $b_x=b$=constant. The estimated $b$ in this case is nothing else but the related parameter of the Weibull model. We can estimate this parameter by fitting a Weibull model to the death probability density function or by the direct estimate from the life tables with the method already discussed and applied in Skiadas and Skiadas (2018). The direct estimate provided $b$=8.1 compared to 8.7 for the estimate via the Weibull model for females in Sweden in 1950. The related figures for 2015 are $b$=10.9 with the direct estimate for b=11.6 via the Weibull model. Note that the figure for the Healthy Life Years Lost provided by the World Health Organization is 10.7 years of age, very close to our direct estimates for $b$ in 2015. Our estimates with both methods (direct and Weibull) are presented in the next Table III along with the WHO estimates for the HLYL. The estimates with the direct method are closer to the WHO. The direct method estimates $b_x$ in all the period of the lifespan thus providing a flexible tool to compare the results provided by WHO at age 60 for females in Sweden. Table III summarizes the related figures. Both methods approach quite well each other. The Direct method estimates $b_x$ for all the life span and we can compare the related results with the WHO findings at 60 years of age presented in Table IV for the years 2000, 2005, 2010, 2015 and 2016. The results verify that both methods approach well between each other. Of course the Direct method, based on only the life tables can used in all the time periods as far as life tables exist.

| TABLE III | | | | | |
|---|---|---|---|---|---|
| Comparisons of *b* estimates with the Healthy Life Years Lost numbers provided by WHO | | | | | |
| Year | 2000 | 2005 | 2010 | 2015 | 2016 |
| WHO | 10.3 | 10.4 | 10.5 | 10.7 | 10.8 |
| Direct method | 10.0 | 10.4 | 10.7 | 10.9 | 10.9 |
| Weibull | 10.7 | 11.0 | 11.4 | 11.6 | 11.6 |

| TABLE IV | | | | | |
|---|---|---|---|---|---|
| Comparisons of $b_x$ estimates with the Healthy Life Years Lost numbers provided by WHO at age 60 | | | | | |
| Year | 2000 | 2005 | 2010 | 2015 | 2016 |
| WHO | 5.3 | 5.4 | 5.6 | 5.7 | 5.8 |
| Direct method | 5.2 | 5.1 | 5.5 | 5.3 | 5.4 |

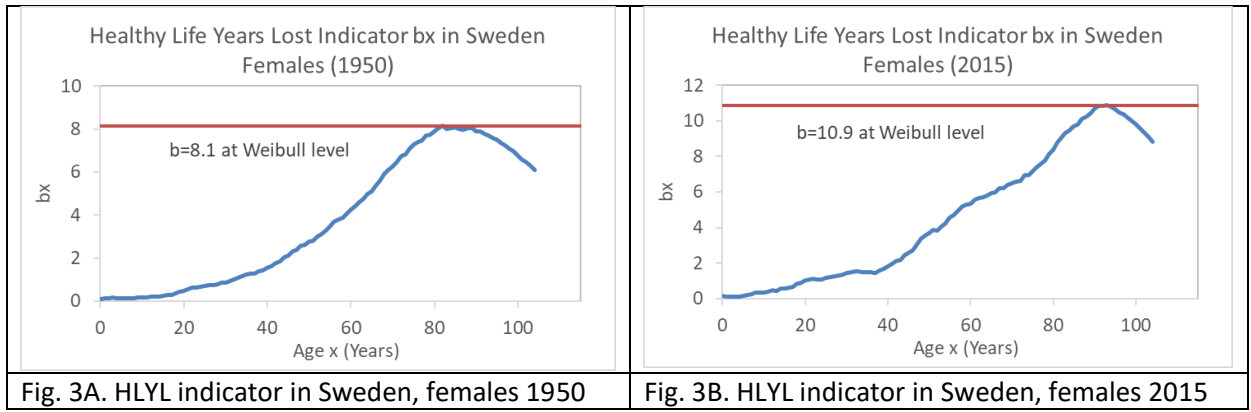

| Fig. 3A. HLYL indicator in Sweden, females 1950 | Fig. 3B. HLYL indicator in Sweden, females 2015 |
|---|---|

The Health/Mortality fraction as presented in figures 3A and 3B has an increasing form for the main part of the lifetime until a maximum and then a decline. In the next figure 4 the case of USA at 2011 is studied. The fraction is similar until 60 years of age with exception of the ages 15 – 30 years when male have an excess of mortality. After 60 years of age female show higher values than male with a maximum of 9.85 at age 96 compared to 9.01 for male at 93 years of age. The very important point here is that the maximum points correspond to years lost to disability. We can easily observe this important future by considering a linear form for mortality mx=ax. This is the simplest case of drawing a linear line from O to B in the graph above. The resulting fraction is 2 whereas is 1 if we select the fraction (ABDOA)/(BCODB). Following the previous discussion, the healthy life years lost to disability (HLYL) are 1 with the last notation and 2 when considering the total space vs the mortality space. The second higher by 1 from the simple fraction provides results similar to those estimated by the World Health Organization. After that, the only we have is to remove this estimate from the life expectancy at birth to find the Healthy Life Expectancy. As it is demonstrated in the graph the HLYL for female are higher than male in the case studied. It is a universal like estimate for the majority of countries.

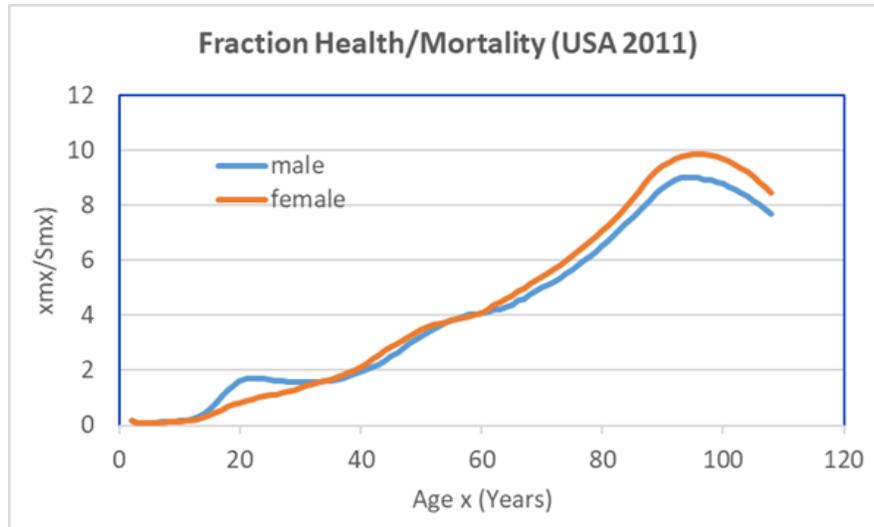

Fig. 4. Fraction Health/Mortality in USA, males and females in 2011

**Program for the Estimates**

We have developed an Excel program for the Direct Estimates of $b_x$ which is provided free of charge. One version can be downloaded from the Demographics 2019 Workshop website at www.asmda.es . The program uses the full life tables from the human mortality database to provide the Healthy Life Year Lost estimator $b_x$ from the general equation form (1):

$$b_x = \frac{x\mu_x}{\int_0^x \mu_s ds}$$

The Cumulative Mortality Mx is given by

$$M_x = \int_0^x \mu_s ds \approx \Sigma_0^x \left(\frac{dx}{lx}\right) \qquad (7)$$

Where dx expresses the death population at age x in the life tables of the HMD and lx is the remaining population at age x in the same life tables. Note that the starting population at age x=0 is set at 100000.

The average mortality Mx/x is estimated by

$$\bar{M}_x = \frac{M_x}{x} \approx \frac{\Sigma_0^x \left(\frac{dx}{lx}\right)}{x}$$

Then the Person Life Years Lost (PLYL) are provided by

$$PLYL = \frac{dx}{\bar{M}_x} = \frac{xdx}{M_x}$$

The final estimate for $b_x$ is given by

$$b_x = \frac{x\mu_x}{\int_0^x \mu_s ds} \approx \frac{PLYL}{lx} = \frac{xdx}{l_x M_x} = \frac{xdx}{l_x \sum_0^x \left(\frac{dx}{lx}\right)} \tag{8}$$

The methodology is presented in the following figure 5. The full life table from the HMD is followed by 4 more columns for the estimation of $b_x$. In the first, the cumulative mortality is estimated from $M = \sum_0^x \mu_x$. The average mortality $(M/x) = \sum_0^x \mu_x / x$ is provided in the next column whereas the Person Life Years Lost (PLYL)=$xd_x/(\sum_0^x \mu_x)$ are calculated in the following column. Where $d_x$ is provided from the column indicated by dx in the life table. For this very important information an interesting graph is provided. The graph follows a growth process until a high level at 77 years of age and a decline in the remaining lifespan period. It the next column the Healthy Life Year Lost estimator $b_x$ is provided by dividing the PLYL with the lx from the life table. The results are presented in an illustrative graph with the growing trend for $b_x$ to reach a maximum at 9.71 with a decline at higher ages. This high level can be also estimated by fitting the Weibull model.

Another option added in this Excel is the estimates of the World Health Organization from 2000-2016 for Life Expectancy at birth and at 60 years of age for all the member countries whereas information for the Healthy Life Expectancy (HALE) at birth and at 60 years of age is provided for the years 2000, 2005, 2010, 2015 and 2016. We have added a small Table to present comparatively the WHO estimates with our estimates with the direct method. The only needed after copy and paste the life table from the HMD to select the name of the country in L1 and the gender (male, female or both sexes) in L2 in the Excel chart. To avoid mistakes we have used list of the WHO countries with their official names.

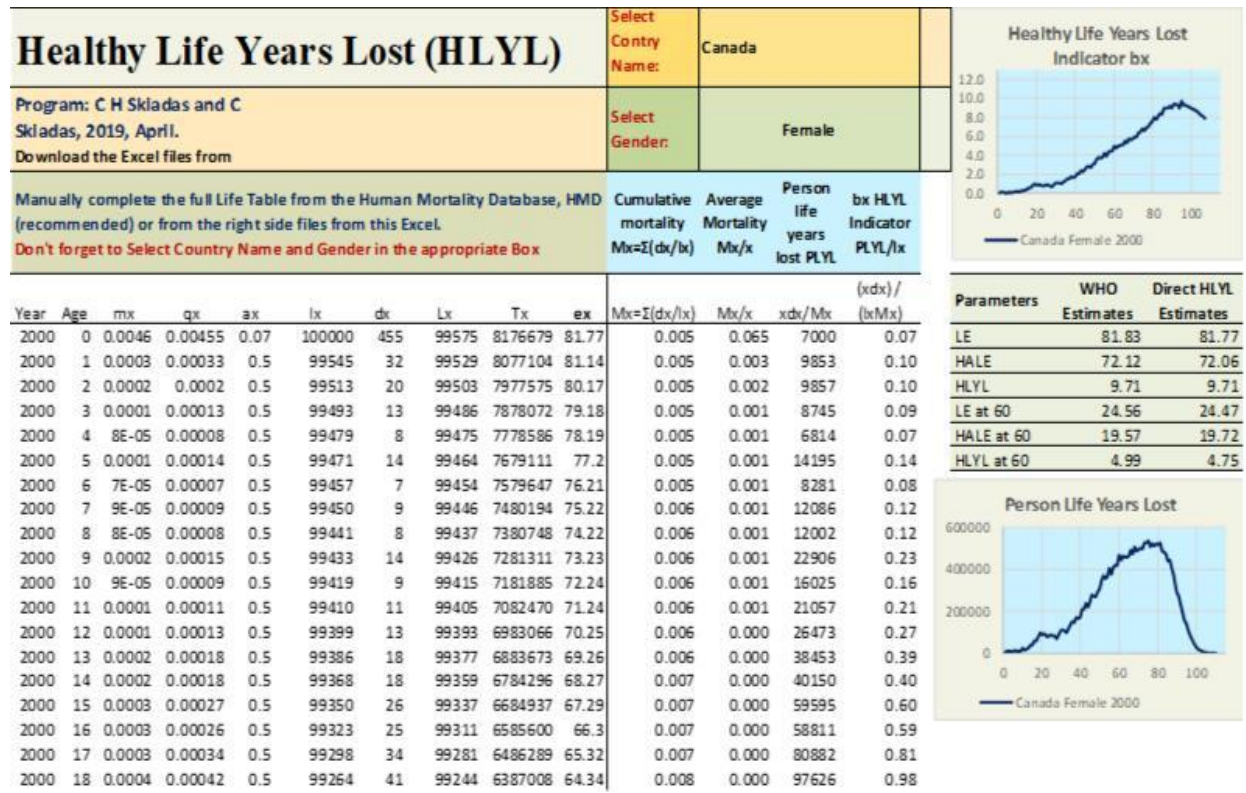

Fig. 5. The extended Life Table for the HLYL estimates

**Conclusions and Further Study**

We have provided an analytical explanation of the behavior of the shape parameter of the Weibull model verifying and expanding the arguments done by Matsushita et al (1992). We have also present an analytic formulation for the observations done along with the development of the appropriate extensions of the classical life tables so that to give a valuable tool for estimating the Healthy Life Years Lost. We have also presented on how the Weibull model properties expressing the fatigue of materials and especially the cumulative hazard of this model can express the additive process of disabilities and diseases to human population. In this paper we further analytically derive a more general model of survival-mortality in which we estimate a parameter related to the Healthy Life Years Lost (HLYL) and leading to the Weibull model and the corresponding shape parameter as a specific case. We have also demonstrated that the results found for the general HLYL parameter we have proposed provides results similar to those provided by the World Health Organization for the Healthy Life Expectancy (HALE) and the corresponding HLYL estimates. An analytic derivation of the mathematical formulas is presented along with an easy to apply Excel program. This program is an extension of the classical life table including four more columns to estimate the cumulative mortality, the average mortality, the person life years lost and finally the HLYL parameter $b_x$. The last versions of this program appear in the Demographics2019 website at: http://www.asmda.es/demographics2019.html .

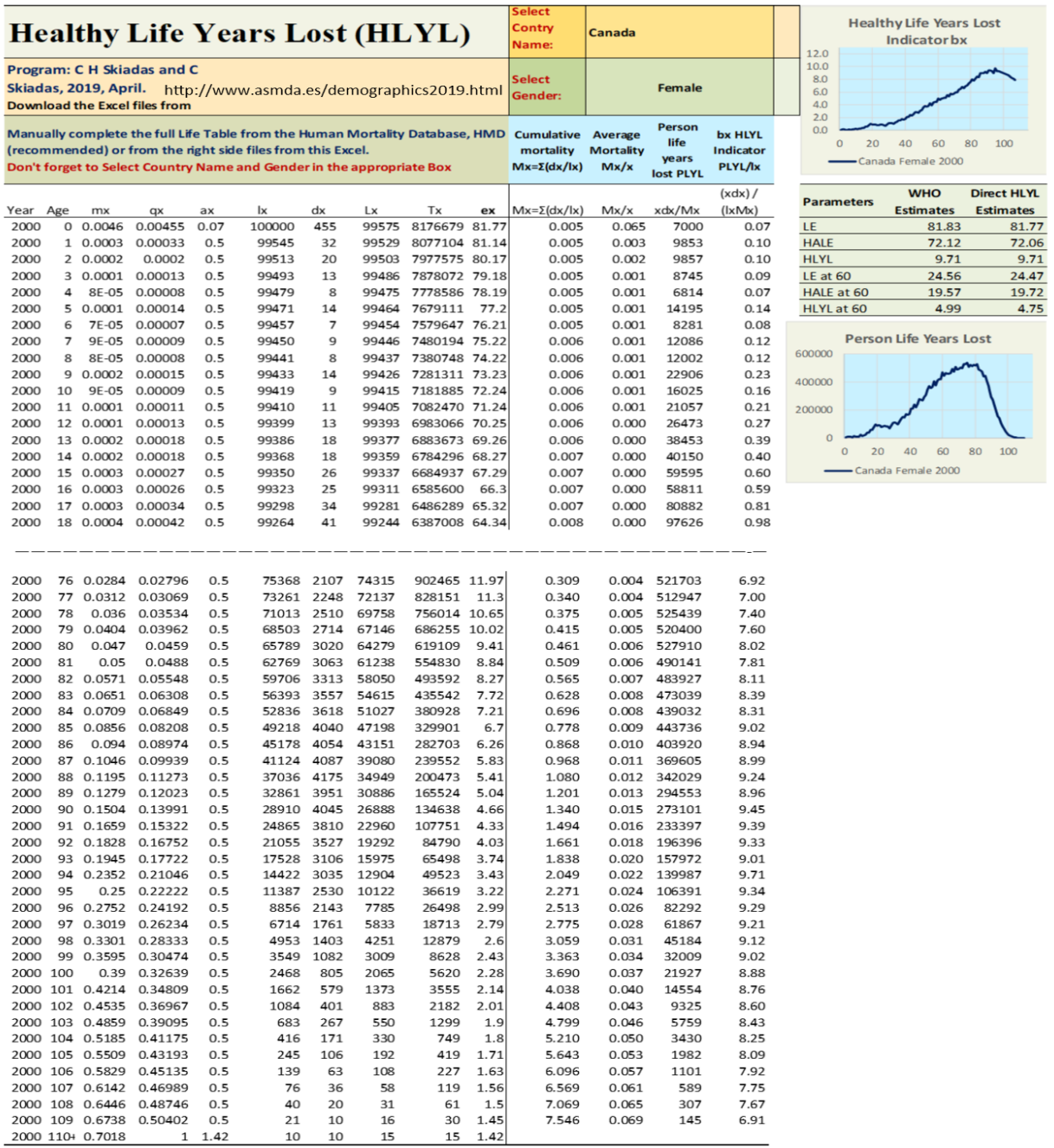

Fig. 6. The extended Life Table for the HLYL estimates. Full presentation